\newcommand{\green}{f_{_{\rm G}}}
\documentclass[psfig]{kluwer}    % Specifies the document style.
\usepackage{graphicx}
\newdisplay{guess}{Conjecture}
\begin{document}                                                                                   
\begin{article}
\begin{opening}         

\title{Further constraints on electron acceleration in solar noise storms}
\author{Prasad \surname{Subramanian}\thanks{e-mail: psubrama@iucaa.ernet.in}}

\institute{Inter-University Centre for Astronomy and Astrophysics,
P.O. Bag 4, Ganeshkhind, Pune - 411007, India}
\author{Peter A. \surname{Becker}\thanks{e-mail: pbecker@gmu.edu}
\thanks{also Dept. of Physics and Astronomy,
George Mason University, Fairfax, VA 22030, USA}}
\institute{Center for Earth Observing and Space Research, \break School of
Computational Sciences, \break George Mason University, Fairfax,
VA 22030, USA}
\runningauthor{Subramanian et al.}
\runningtitle{Electron acceleration in solar noise storms}

%\date{}

\begin{abstract}
We reexamine the energetics of nonthermal electron acceleration in solar
noise storms. A new result is obtained for the minimum nonthermal
electron number density required to produce a Langmuir wave population
of sufficient intensity to power the noise storm emission. We combine
this constraint with the stochastic electron acceleration formalism
developed by Subramanian \& Becker (2005) to derive a rigorous estimate
for the efficiency of the overall noise storm emission process,
beginning with nonthermal electron acceleration and culminating in the
observed radiation. We also calculate separate efficiencies for the
electron acceleration -- Langmuir wave generation stage and the Langmuir
wave -- noise storm production stage. In addition, we obtain a new
theoretical estimate for the energy density of the Langmuir waves in
noise storm continuum sources.
\end{abstract}

\end{opening}

\section{Introduction}

Noise storm radio emission from the solar corona is typically manifested
as long-lived, broadband noise (continuum) radiation at
metric/decimetric wavelengths, with occasional short-lived (0.1--1~s),
intense, narrowband bursts. The storms are typically observed in the
50-500~MHz range, and are brightest around 150-200~MHz. At 164~MHz, the
peak intensity of noise storm continua is $\sim 20$ SFU while that of
the short bursts is $\sim 100$ SFU (Malik \& Mercier 1996; Kerdraon
1973; Kerdraon 1979).

\subsection{Estimate of Total Power}

Aside from solar flares, noise storms are the only known sites of
electron acceleration in the corona. Estimates of the power budget
involved in the electron acceleration process are few and sketchy, and
this will be the focus of our paper. We first turn our attention to the
power emitted in noise storm radiation. Elgar\o y (1977) used the
oft-quoted multiwavelength observational study of 28 noise storms
conducted by Smerd (1964) to arrive at a figure of $ F_{\rm out} \sim 3
\times 10^{-12}\,{\rm W}\,{\rm m}^{-2}$ for the frequency-integrated
energy flux measured at the earth for a typical noise storm. He then
estimated the total noise storm power, $L_{\rm out}$, by writing
\begin{equation}
L_{\rm out} = F_{\rm out}\,R^{2}\,\Omega\,e^{-\tau} \ ,
\label{neweq001}
\end{equation}
where $R$ is the sun-earth distance, $\Omega$ is the solid angle through
which the radiation is beamed, and $\tau$ is the optical depth of the
source. By setting $\tau = 2$ and $\Omega = 0.6$~ster, Elgar\o y obtained
the canonical result
\begin{equation}
L_{\rm out} \sim 10^{17}\,{\rm erg}\,{\rm s}^{-1} \ .
\label{neweq002}
\end{equation}
%
%$L_{\rm out} \sim 10^{17}\,{\rm erg}\,{\rm s}^{-1}$.

Solar noise storm radiation is believed to form via a two-stage process.
First, thermal electrons are accelerated to form a nonthermal electron
population, and second, the Langmuir waves produced by the nonthermal
electrons are converted into observable radio emission via coalescence
with a suitable population of low-frequency waves. Melrose (1975),
Robinson (1978), and Wentzel (1985) have considered the issue of
Langmuir wave generation by a population of electrons. While any
electron population will generate some level of Langmuir waves via
spontaneous emission, the production of bright noise storms requires an
appreciable power in the Langmuir waves.

\subsection{Review of Previous Results}

We have analyzed in a previous paper the evolution of the Green's
function $\green$ describing the nonthermal electron distribution
resulting from the stochastic acceleration of particles injected with a
monoenergetic distribution (Subramanian \& Becker, 2005, hereafter
paper~1). In a steady-state situation, the Green's function is governed
by a linear, partial differential Fokker-Planck transport equation with
momentum diffusion coefficient ${\cal D}$ given by
\begin{equation}
{\cal D} = D_0 \, p^2 \ ,
\label{neweq1}
\end{equation}
where $D_0$ is a constant with the units of inverse time. This specific
form for ${\cal D}$ is motivated by a number of particle transport
scenarios such as the acceleration of electrons by large-scale
compressible magnetohydrodynamical (MHD) turbulence (Ptuskin, 1988;
Chandran \& Maron, 2003); the energization of electrons by cascading
fast-mode waves in flares (Miller, LaRosa \& Moore, 1996); and the
acceleration of electrons due to lower hybrid turbulence (Luo et al.,
2003).

The resulting Green's function for the nonthermal electron
distribution is found to be (see paper~1; also Subramanian, Becker,
\& Kazanas, 1999)
\begin{equation}
\green(p,p_0) = A_0 \, \cases{
(p/p_0)^{\alpha_1} \ , & $p \le p_0$ \ , \cr
\phantom{space} \cr
(p/p_0)^{\alpha_2} \ , & $p \ge p_0$ \ , \cr 
}
\label{neweq2}
\end{equation}
where $p$ is the electron momentum, $p_0$ is the momentum of the
injected monoenergetic electrons, and the exponents $\alpha_1$ and
$\alpha_2$ are given by
\begin{equation}
\alpha_1 \equiv - {3 \over 2} + \left({9 \over 4}
+ {1 \over D_0 \, \tau}\right)^{1/2} \ , \ \ \ \ \
\alpha_2 \equiv - {3 \over 2} - \left({9 \over 4}
+ {1 \over D_0 \, \tau} \right)^{1/2} \ .
\label{neweq3}
\end{equation}
The quantity $\tau$ in these expressions represents the mean residence
time for electrons in the acceleration region, and the normalization
parameter $A_0$ is computed using
\begin{equation}
A_0 \equiv {\dot N_0 \over 2 \, D_0 \, p_0^3} \left({9 \over 4} +
{1 \over D_0 \, \tau}\right)^{-1/2} \ ,
\label{neweq4}
\end{equation}
where the constant $\dot N_0$ denotes the number of electrons injected
per unit time per unit volume into the acceleration region. The value of
the total electron number density associated with the Green's function
distribution is given by
\begin{equation}
n_{_{\rm G}} \equiv \int_0^\infty p^2 \, \green  \, dp = \dot N_0 \, \tau \ ,
\label{neweq5}
\end{equation}
as expected in this steady-state situation. Since the nonthermal
electrons are nonrelativistic with kinetic energy $p^2/(2 \, m_e)$, it
follows that we must require $\alpha_2 < -5$ in order to avoid an
infinite energy density, and therefore $D_0 \, \tau < 10^{-1}$.

In paper~1, we applied the formalism described above to model the
transport in momentum space of electrons injected from the high-energy
portion of the Maxwellian distribution in the corona. We found that
stochastic acceleration of the electrons dominates over losses due
to collisions and Langmuir damping for particles with momenta
$p > p_c$, where
\begin{equation}
p_c \equiv 5.35 \times 10^{-22} \left(\Lambda \, n_e \over D_0
\right)^{1/3}
\label{neweq6}
\end{equation}
denotes the ``critical momentum'' in cgs units, $\Lambda$ is the Coulomb
logarithm, and $n_e$ represents the total electron number density in the
corona. Electrons with $p > p_c$ experience net acceleration on average,
and those with $p < p_c$ are decelerated on average. It is expected that
a ``gap'' distribution will form as a result of the collisional and
Langmuir losses experienced by electrons with $p < p_c$.

Based on analysis of a generic second-order Fermi (stochastic)
acceleration mechanism, we demonstrated in paper~1 that if all of the
electrons in the Maxwellian distribution with $p > p_c$ are subject to
acceleration, then the nonthermal electron fraction is given by
\begin{eqnarray}
&{n_* \over n_e} =
{2 \, \xi_c^{(3+\alpha_1)/2} \, \Gamma\left(- {\alpha_1 \over 2}
, \, \xi_c \right) \over \sqrt{\pi} \, (3+\alpha_1) (\alpha_2-\alpha_1)
\, D_0 \, \tau}
\nonumber \\
&\phantom{lotsofspaaace} + \, 2  \, e^{-\xi_c} \left(\xi_c \over
\pi\right)^{1/2} + {\rm Erfc}\left(\xi_c^{1/2}\right)
\ ,
\label{neweq16}
\end{eqnarray}
where $n_*$ denotes the nonthermal electron number density, $\alpha_1$
and $\alpha_2$ are the power-law indices given by Equations~(\ref{neweq3}),
and the dimensionless critical electron energy, $\xi_c$, is defined by
\begin{equation}
\xi_c \equiv {p_c^2 \over 2 \, m_e k_{\rm B} T_e} \ ,
\label{neweq17}
\end{equation}
with $T_e$ denoting the temperature of the thermal electrons, $m_e$ the
electron mass, and $k_{\rm B}$ Boltzmann's constant.

The quantities $\alpha_1$ and $D_0 \, \tau$ can be expressed in
terms of the high-energy power-law index $\alpha_2$ by using
Equations~(\ref{neweq3}) to write
\begin{equation}
\alpha_1 = - 3 - \alpha_2 \ ,
\ \ \ \ \ 
D_0 \, \tau = {1 \over \alpha_2 \, (3 + \alpha_2)}
\ .
\label{neweq18}
\end{equation}
These relations can be combined with Equation~(\ref{neweq16})
to obtain the alternative result for the nonthermal electron
fraction
\begin{eqnarray}
&{n_* \over n_e} =
- {2 \, \xi_c^{-\alpha_2/2} \, (3+\alpha_2) \,
\Gamma\left({3+\alpha_2 \over 2}
, \, \xi_c \right) \over \sqrt{\pi} \ (3+2\alpha_2)}
\nonumber \\
&\phantom{lotsofspaaace} + \, 2  \, e^{-\xi_c}
\left(\xi_c \over \pi\right)^{1/2}
+ {\rm Erfc}\left(\xi_c^{1/2}\right)
\ ,
\label{neweq19}
\end{eqnarray}
which is convenient because the right-hand side is now an explicit
function of $\alpha_2$ and $\xi_c$. In paper~1, both $\alpha_2$ and
$n_*/n_e$ were treated as free parameters, but in the present
paper the second quantity is computed self-consistently using
equation~(\ref{neweq19}). Hence the only remaining free parameter
is $\alpha_2$.

\section{Langmuir wave generation and the ``gap'' electron distribution}

For the sake of simplicity, we concentrate here on isotropic electron
distributions, in which case the most suitable candidate for producing a
sufficiently intense population of Langmuir waves is a gap electron
distribution (Melrose, 1975). Such a gap distribution is often idealized
as a narrow ``hump'' of nonthermal electrons located far above the
thermal distribution in the energy space. Melrose (1975) elaborates on
the conditions for the formation of a gap distribution and on the
parameters of such a distribution required to ensure a sufficiently
intense population of Langmuir waves. The gap is the result of
collisional and Langmuir losses experienced by particles in the energy
range between the typical thermal energy and highly superthermal
energies. In a steady state, the intensity of the Langmuir waves is
governed by the balance between emission and absorption of the waves by
the thermal and nonthermal electrons. A gap electron distribution with
the right parameters allows superthermal electrons to emit Langmuir
waves without reabsorbing them, resulting in a high wave intensity.

We now turn our attention to the specific parameters of a gap electron
distribution that will enable it to produce a high intensity Langmuir
wave population, which, in turn, is needed to produce bright noise storm
emission. Consider a gap distribution where the nonthermal electrons are
peaked around the critical momentum, $p_c$, given by
Equation~(\ref{neweq6}). In this case, Melrose (1975) points out that in
order to form an intense Langmuir wave distribution, the number density
of the nonthermal electrons must be high enough for them to dominate
over the thermal electrons in emitting and absorbing waves with phase
velocities $v_\phi \approx v_c$, where $v_c \equiv p_c/m_e$ is the
``critical velocity.'' The phase velocity
of the Langmuir waves is related to the wavenumber $k$ via
\begin{equation}
v_\phi = {\omega_p \over k} \ ,
\label{neweq7}
\end{equation}
where
\begin{equation}
\omega_p = \left(4 \pi e^2 n_e \over m_e\right)^{1/2}
\label{neweq8}
\end{equation}
denotes the electron plasma frequency and $e$ is the electron charge.
The formation of a high intensity (or, equivalently, a high brightness
temperature) Langmuir wave population requires that the wave damping due
to the nonthermal electrons must dominate over that due to the thermal
electrons for waves with wavenumber $k\approx k_c$, where
\begin{equation}
k_c \equiv {\omega_p \over v_c}
= {\omega_p \, m_e \over p_c} \ .
\label{neweq9}
\end{equation}
In the following analysis, we will concentrate on the constraints
imposed by this criterion on the physical mechanism that accelerates the
nonthermal electrons with $p \ge p_c$.

Based on Equation~(20) from Robinson (1978), we write the collisional
damping rate for the Langmuir waves in cgs units as
\begin{equation}
\gamma_0 = 30 \, {n_{\rm th} \over T_e^{3/2}}
\ \propto \ {\rm s^{-1}} \ ,
\label{neweq10}
\end{equation}
where $n_{\rm th}$ is the number density of the thermal electrons,
and the value of
$\gamma_0$ is independent of the wavenumber $k$. Note that the thermal
electron number density $n_{\rm th}$ is essentially equal to the total
electron number density $n_e$ appearing in Equation~(\ref{neweq6})
because only the small population of electrons with $p > p_c$ in the
tail of the Maxwellian experiences acceleration. The wave damping rate
due to interactions with the nonthermal electrons in the gap
distribution is given by Equation~(18) from Robinson (1978), which
states that
\begin{equation}
\gamma_{\rm gap}(k) = {8 \pi^2 n_* \, e^2 \omega_l^2(k)
\over k^3 c^2 p_c} \ ,
\label{neweq11}
\end{equation}
where $n_* \ll n_{\rm th}$ is the number density of the nonthermal
electrons, and the Langmuir frequency, $\omega_l(k)$, is given by
the dispersion relation (e.g., Robinson, 1978)
\begin{equation}
\omega_l^2(k) = \omega_p^2 + 3 k^2 v_{\rm th}^2
\label{neweq12}
\end{equation}
for electrons with characteristic thermal velocity $v_{\rm th} \equiv
(k_{\rm B} T_e/m_e)^{1/2}$.

The requirement that the nonthermal electrons dominate over the thermal
particles in damping the Langmuir waves with $k \approx k_c$ can be
stated quantitatively as
\begin{equation}
\gamma_{\rm gap}(k_c) \ > \ \gamma_0
\ ,
\label{neweq13}
\end{equation}
or, equivalently,
\begin{equation}
{n_* \over n_e} > 1.5 \times 10^{39} \, {k_c^3 \, p_c
\over T_e^{3/2} \omega_l^2(k_c)}
\ ,
\label{neweq14}
\end{equation}
in cgs units, where we have employed Equations~(\ref{neweq10}) and
(\ref{neweq11}) along with the fact that $n_{\rm th} \approx n_e$ to
obtain the final result. Equation~(\ref{neweq14}) represents an
interesting constraint on the nonthermal electron fraction $n_*/n_e$ in
the solar corona during the production of the noise storm emission.
The right-hand side of Equation~(\ref{neweq14}) can be stated in terms
of $p_c$, $T_e$, and $\omega_p$ by employing Equations~(\ref{neweq9})
and (\ref{neweq12}), which yields
\begin{equation}
{n_* \over n_e} > 1.1 \times 10^{-42} \, {\omega_p \over
T_e^{3/2} (p_c^2 + 3 k_{\rm B} T_e \, m_e)} \ .
\label{neweq15}
\end{equation}
Satisfaction of this inequality ensures that the intensity of the
Langmuir waves generated by the gap distribution is sufficient to power
the observed noise storm emission.

\section{Acceleration of nonthermal electrons}

We can use Equation~(\ref{neweq17}) to reexpress the right-hand side of
Equation~(\ref{neweq15}) in terms of $\xi_c$, $\omega_p$, and $T_e$,
which yields
\begin{equation}
{n_* \over n_e} > {8.9 \, \omega_p \over
T_e^{5/2} (2 \, \xi_c + 3)}
\label{neweq20}
\end{equation}
in cgs units. The nonthermal density fraction must satisfy this
inequality in order to guarantee a sufficiently intense Langmuir wave
distribution. By substituting for the nonthermal electron fraction
$n_*/n_e$ in Equation~(\ref{neweq20}) using Equation~(\ref{neweq19}), we
can show that the dimensionless critical energy $\xi_c$ must satisfy the
corresponding inequality

\def\ximax{\xi_{\rm max}}

\begin{equation}
\xi_c < \ximax \ ,
\label{neweq21}
\end{equation}
where $\ximax$ is defined by the relation
\begin{eqnarray}
&{4.4\,\omega_p \over T_e^{5/2}\,(\ximax + 1.5)} =
- {2 \, \ximax^{-\alpha_2/2} \, (3+\alpha_2) \,
\Gamma\left({3+\alpha_2 \over 2}
, \, \ximax \right) \over \sqrt{\pi} \ (3+2\alpha_2)}
\nonumber \\
&\phantom{lotsofspaaace} + \, 2  \, e^{-\ximax}
\left(\ximax \over \pi\right)^{1/2}
+ {\rm Erfc}\left(\ximax^{1/2}\right)
\ .
\label{neweq22}
\end{eqnarray}
The quantity $\ximax$, which is a function of $T_e$, $\omega_p$, and
$\alpha_2$, represents the {\it maximum value} of $\xi_c$ such that the
intensity of the associated Langmuir wave distribution is sufficient to
power the observed noise storm emission.

\begin{figure}
\includegraphics[width=5.5in]{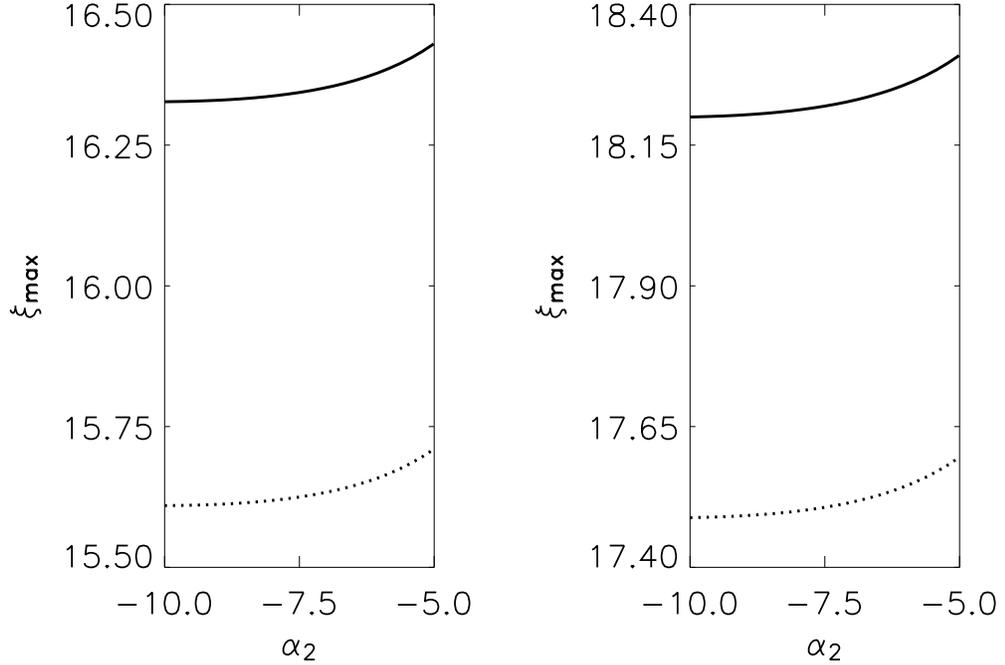}
\label{Fig 1}
\caption{Maximum value of the dimensionless critical energy, $\ximax$,
plotted as a function of $\alpha_2$ using Equation~(\ref{neweq22}). Left
panel: $T_e = 10^6\,$K. Right panel: $T_e = 2 \times 10^6\,$K. Solid
line: $\nu = 169\,$MHz. Dotted line: $\nu = 327\,$MHz.}
\end{figure}

In Figure~1 we plot $\ximax$ as a function of $\alpha_2$ for observing
frequencies $\nu \equiv \omega_p/(2 \pi) = 169\,$MHz and $327\,$MHz and
for coronal temperatures $T_e = 10^6\,$K and $2 \times 10^6\,$K. 
Figure~1 indicates that $\ximax$ has a moderate dependence on the
observing frequency $\nu$ and the coronal temperature $T_e$, but it is
quite insensitive to the value of $\alpha_2$.

\begin{figure}
\includegraphics[width=5.5in]{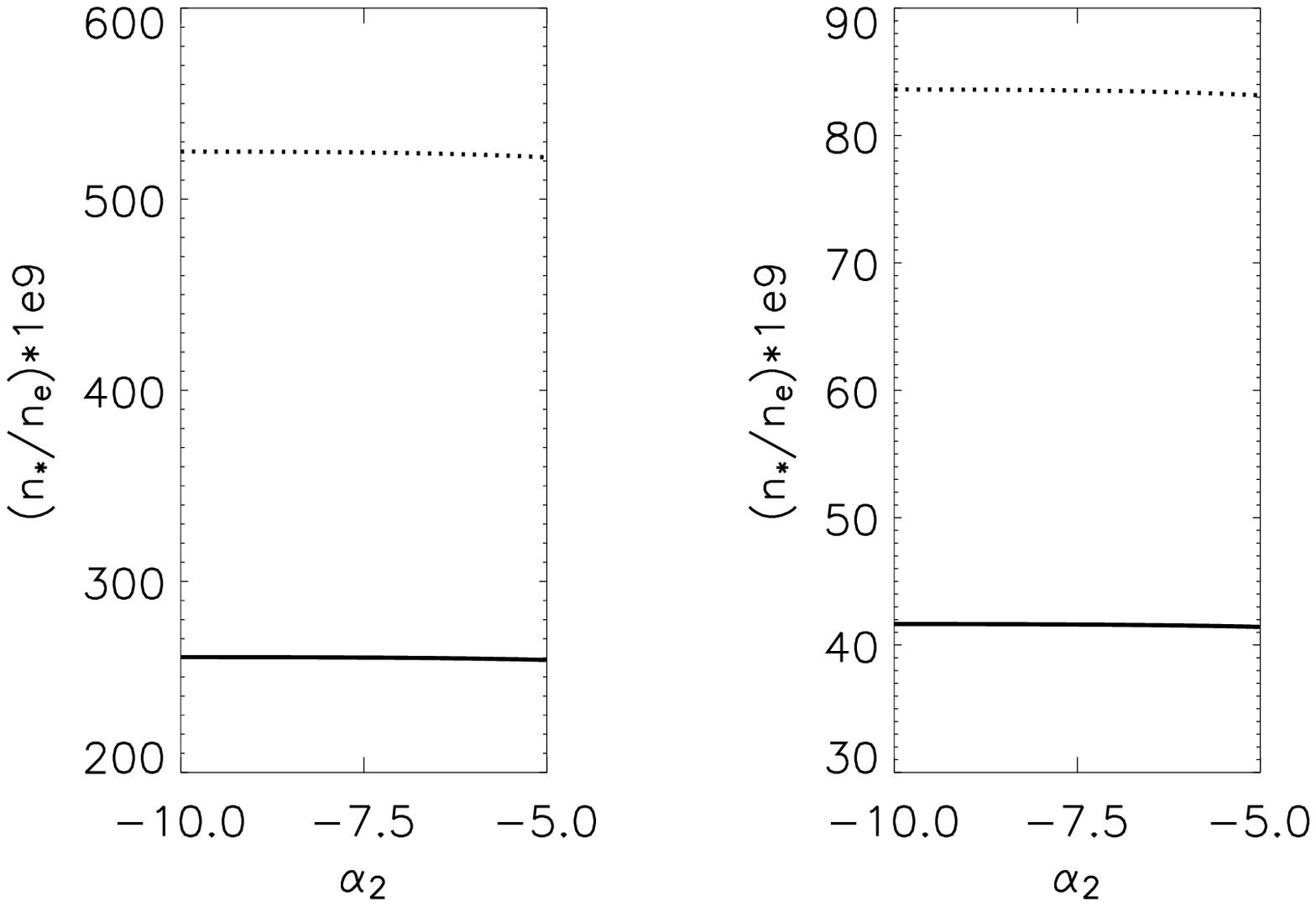}
\label{Fig 2}
\caption{Minimum nonthermal electron fraction $n_*/n_e$ plotted as a
function of $\alpha_2$ using Equation~(\ref{neweq19}) with $\xi_c =
\ximax$. Left panel: $T_e = 10^6\,$K. Right panel: $T_e = 2 \times
10^6\,$K. Solid line: $\nu = 169\,$MHz. Dotted line: $\nu = 327\,$MHz.}
\end{figure}

In Figure~2 we plot the nonthermal electron fraction $n_*/n_e$ evaluated
using Equation~(\ref{neweq19}) with $\xi_c = \ximax$ as a function of
the high-energy power-law index $\alpha_2$ based on the same values for
the observing frequency $\nu$ and coronal temperature $T_e$ used in
Figure~1. Since we have set $\xi_c = \ximax$, it follows that the curves
in Figure~2 represent the {\it minimum} values for $n_*/n_e$ such that
the Langmuir wave distribution has sufficient intensity to produce the
observed noise storm emission. This ratio exhibits marked dependences on
$\nu$ and $T_e$, but it is clearly a very weak function of $\alpha_2$.
We note that for a coronal electron temperature $T_e = 10^6\,$K and an
observing frequency $\nu = 169\,$MHz, the nonthermal electron fraction
indicated by Figure~2 is close to the value $n_* /n_e = 2.2 \times
10^{-7}$ which Thejappa (1991) demonstrated to be the threshold value
for the transition between the production of steady noise storm emission
and the development of intense type~I bursts. Based on Thejappa's
result, the value $n_* /n_e = 2.2 \times 10^{-7}$ was adopted as an ad
hoc input parameter in paper~1, but in the present paper, this ratio is
calculated self-consistently based on the physics of the electron
acceleration -- Langmuir wave generation processes. We provide further
discussion of the relationship between our model and that of Thejappa in
Section~6.

\section{Nonthermal electron input power}

In paper~1, we demonstrated that the energy density of the accelerated
particles is given by
\begin{eqnarray}
& {U_* \over n_e k_{\rm B} T_e} =
{2 \, \xi_c^{(5+\alpha_1)/2} \, \Gamma\left(- {\alpha_1 \over 2} ,
\, \xi_c \right) \over \sqrt{\pi} \, (5+\alpha_1) (\alpha_2-\alpha_1)
\, D_0 \tau}
\nonumber \\
& \phantom{lotsofspaaace} + \, {2 \sqrt{\pi \xi_c} \, (3 + 2 \,
\xi_c) \, e^{-\xi_c} + \, 3 \pi \, {\rm Erfc}\left(\xi_c^{1/2}
\right) \over 2 \pi (1 - 10 \, D_0 \, \tau)} \ .
\label{neweq23}
\end{eqnarray}
By utilizing Equations~(\ref{neweq18}) to eliminate $\alpha_1$ and
$D_0 \tau$, we can obtain the alternative form
\begin{eqnarray}
& {U_* \over n_e k_{\rm B} T_e} = \alpha_2 (3+\alpha_2)
\bigg[{2 \, \xi_c^{(2-\alpha_2)/2} \, \Gamma\left({3+\alpha_2 \over 2}
, \, \xi_c \right) \over \sqrt{\pi} \, (2-\alpha_2) (3+2\alpha_2)}
\nonumber \\
& \phantom{lotsofspaaace} + \, {2 \sqrt{\pi \xi_c} \, (3 + 2 \, \xi_c)
\, e^{-\xi_c} + \, 3 \pi \, {\rm Erfc}\left(\xi_c^{1/2}\right)
\over 2 \pi (\alpha_2^2 + 3 \alpha_2 - 10)}\bigg] \ ,
\label{neweq24}
\end{eqnarray}
where the right-hand side is now an explicit function of $\alpha_2$ and
$\xi_c$.

In Figure~3 we plot the energy density ratio $U_*/(n_e k_{\rm B} T_e)$
as a function of $\alpha_2$ and $T_e$ evaluated using
Equation~(\ref{neweq24}) with $\xi_c = \ximax$. The curves therefore
represent the {\it minimum} values for $U_*/(n_e k_{\rm B} T_e)$ that
are consistent with the required intensity of the Langmuir waves. In
contrast with the nonthermal electron fraction $n_*/n_e$ plotted in
Figure~2, we note that the energy density ratio displays a marked
dependence on the high-energy power-law index $\alpha_2$, diverging as
$\alpha_2 \to -5$ as expected.

\begin{figure}
\includegraphics[width=5.5in]{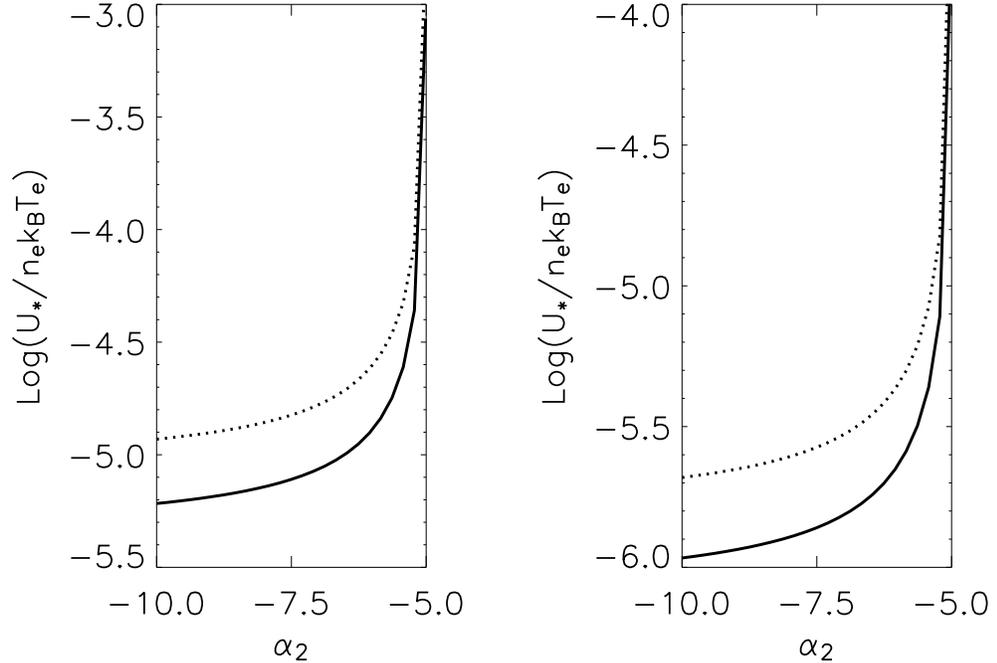}
\label{Fig 3}
\caption{Minimum nonthermal energy density ratio $U_*/(n_e k_{\rm B}
T_e)$ plotted as a function of $\alpha_2$ using Equation~(\ref{neweq24})
with $\xi_c = \ximax$. Left panel: $T_e = 10^6\,$K. Right panel: $T_e =
2 \times 10^6\,$K. Solid line: $\nu = 169\,$MHz. Dotted line: $\nu =
327\,$MHz.}
\end{figure}

Based on Equations~(23) and (29) from paper~1, we can express the total
power required to drive the acceleration of the nonthermal electrons as
\begin{equation}
L_{\rm in} = 8 V D_0 \, U_* \ ,
\label{neweq25}
\end{equation}
where $U_*$ is given by Equation~(\ref{neweq24}), $V$ is the volume of
the acceleration/emission region, and the diffusion constant $D_0$ can
be evaluated in terms of $\xi_c$ using (see Equations~(\ref{neweq6}) and
(\ref{neweq17}))
\begin{equation}
D_0 = {1.2 \, \Lambda \, n_e \over \xi_c^{3/2} \, T_e^{3/2}}
\ .
\label{neweq26}
\end{equation}

By combining our various relations, the required input power $L_{\rm
in}$ can be evaluated in terms of $\alpha_2$, $\xi_c$, $n_e$, $T_e$,
$\Lambda$, and $V$. In general, we set the Coulomb logarithm $\Lambda =
29.1$ (e.g., Brown, 1972). We find that by setting $\xi_c$ equal to the
maximum value $\ximax$ given by Equation~(\ref{neweq22}), we obtain the
{\it minimum} value for the input power $L_{\rm in}$ consistent with the
constraint that the intensity of the generated Langmuir waves be
sufficient to power the observed noise storm emission. Hence by setting
$\xi_c = \ximax$ and selecting specific values for the physical
parameters $n_e$, $T_e$, and $V$, we can use Equation~(\ref{neweq25}) to
compute the minimum input power $L_{\rm in}$ as a function of only one
parameter, $\alpha_2$.

\section{Results}

\subsection{Efficiency of overall electron acceleration-noise storm
continuum radiation process}

Once the upper limit for the dimensionless critical electron energy,
$\ximax$, has been computed using Equation~(\ref{neweq22}), we can use
Equation~(\ref{neweq25}) to obtain the minimum value for the input power
$L_{\rm in}$ consistent with the required intensity of the Langmuir
waves. The efficiency of the overall noise storm emission process,
beginning with the acceleration of the nonthermal electrons and ending
with the production of the observed radiation, is characterized by the
efficiency parameter $\eta$, defined by
\begin{equation}
\eta \equiv {L_{\rm out} \over L_{\rm in}} \ ,
\label{neweq27}
\end{equation}
where $L_{\rm out}$ denotes the observed power of the noise storm
emission. Since we use the minimum possible value of $L_{\rm in}$, it follows
that the we are in fact computing the {\it maximum value} of the efficiency
$\eta$. The crucial difference between the approach taken here and that
used in paper~1 is that in the present treatment we set $\xi_c = \ximax$,
which allows us to obtain the final result for the overall efficiency $\eta$
in terms of the single parameter $\alpha_2$.

The maximum value of the overall efficiency $\eta$ is plotted in
Figure~4 as a function of $\alpha_2$ for the same two values of the
coronal electron temperature $T_e$ and observing frequency $\nu$ used in
Figure~1. We set the output power and the emission region volume using
$L_{\rm out} = 10^{17}\,{\rm erg \, s}^{-1}$ (see Equation~(\ref{neweq002}))
and $V = 10^{30}\,{\rm cm}^3$, respectively. The total electron number
density $n_e$ is calculated by setting $\omega_p = 2 \pi \nu$ and then
employing Equation~(\ref{neweq8}). In general, it is seen that the
efficiency of the overall process is $\eta \lesssim 10^{-6}$.

\begin{figure}
\includegraphics[width=5.5in]{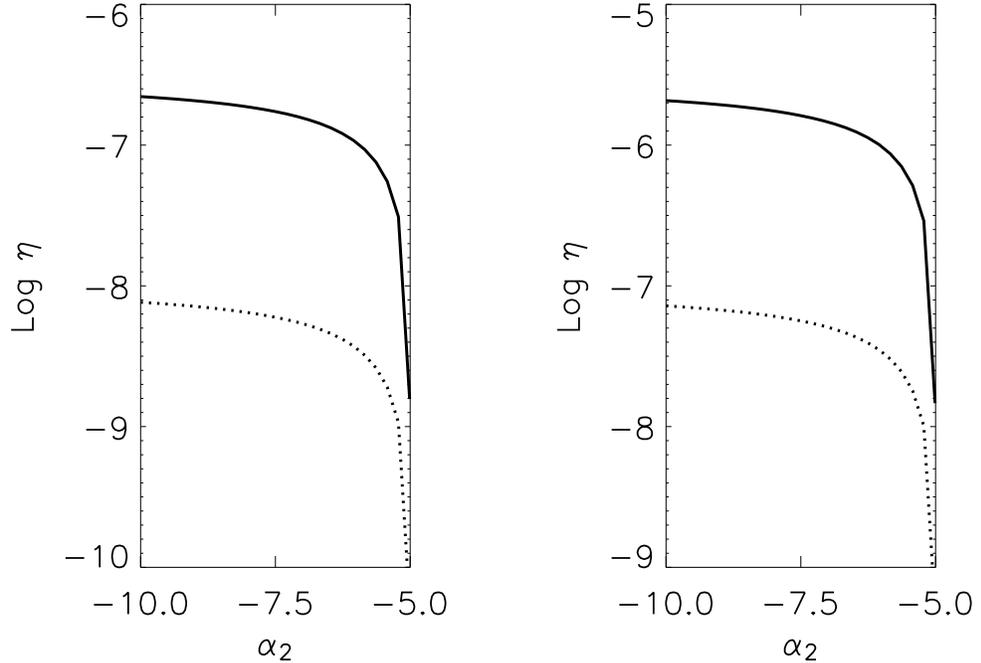}
\label{Fig 4}
\caption{Maximum efficiency of the overall noise storm emission process,
$\eta \equiv L_{\rm out}/L_{\rm in}$, plotted as a function of
$\alpha_2$ using Equation~(\ref{neweq27}) with $\xi_c = \ximax$. Left
panel: $T_e = 10^6\,$K. Right panel: $T_e = 2 \times 10^6\,$K. Solid
line: $\nu = 169\,$MHz. Dotted line: $\nu = 327\,$MHz. The output power
$L_{\rm out}$ is taken to be $10^{17}\,{\rm erg\,s}^{-1}$.}
\end{figure}

\subsection{Efficiency of first stage: electron acceleration-Langmuir
wave generation}

Having computed the efficiency of the overall emission process in
Section~5.1, we next seek to evaluate the efficiency of the first stage
of the scenario by calculating the ratio of the power in the Langmuir
wave population, $L_{\rm L}$, to the power input to the electron
acceleration process, $L_{\rm in}$. The former quantity can be
calculated using (Melrose 1975)
\begin{equation}
L_{\rm L} = {V \over 2} \, {W_{\rm L} \over t_{\rm L}} \ ,
\label{neweq28}
\end{equation}
where $V$ is the volume of the acceleration region, $W_{\rm L}$ is the
energy density in the Langmuir waves, and $t_{\rm L}$ is the Langmuir
wave generation timescale. We can compute $W_{\rm L}$ using
Equation~(20) from Melrose (1975), which gives
\begin{equation}
W_{\rm L} = {n_* m_e v_c^2 \over 2} \,
\biggl[\biggl({v_{\rm th} \over v_c}\biggr)^3 \, \ln
\biggl({v_c \over v_{\rm th}}\biggr)\biggr] \ ,
\label{neweq29}
\end{equation}
where $v_{\rm th} \equiv (k_{\rm B} T_e/m_e)^{1/2}$ is the
characteristic velocity of the thermal electrons, $v_c=p_c/m_e$ is the
critical velocity (the speed of the nonthermal electrons in the gap
distribution), and $n_*$ is the nonthermal electron number density. In
our current application, we set $\xi_c = \ximax$ and compute $n_*$ using
Equation~(\ref{neweq19}). The critical velocity is likewise given by
$v_c = (2 k_{\rm B} T_e \ximax/m_e)^{1/2}$ which follows from
Equation~(\ref{neweq17}).

As discussed in paper~1 and also by Melrose (1980), we assume that the
Langmuir wave generation process is proceeding at marginal stability, in
which case the Langmuir wave generation timescale $t_{\rm L}$ is
comparable to the Coulomb loss timescale $t_{\rm loss}$. We are focusing
here on Langmuir waves with wavenumber $k \approx k_c$ (see
Equation~(\ref{neweq9})), and therefore the relevant Coulomb loss
timescale is associated with electrons with momentum $p \approx p_c$. It
follows that we can evaluate the Langmuir wave generation timescale
$t_{\rm L} \approx t_{\rm loss}$ using Equation~(4) of paper~1 with
$\epsilon = p_c^2/(2 m_e)$, which can be written in terms of the
dimensionless critical energy $\xi_c$ as
\begin{equation}
t_{\rm L} = 0.10 \ {\xi_c^{3/2} \, T_e^{3/2} \over \Lambda \, n_e}
\label{neweq30}
\end{equation}
in cgs units. We adopt the value $\Lambda = 29.1$ in our numerical
calculations, and set $\xi_c = \ximax$ in order to treat the critical
case of interest here. Detailed analysis establishes that the result
obtained for $L_{\rm L}$ using Equations~(\ref{neweq28}),
(\ref{neweq29}), and (\ref{neweq30}) is the {\it minimum} value
consistent with the required Langmuir wave intensity.

\begin{figure}
\includegraphics[width=5.5in]{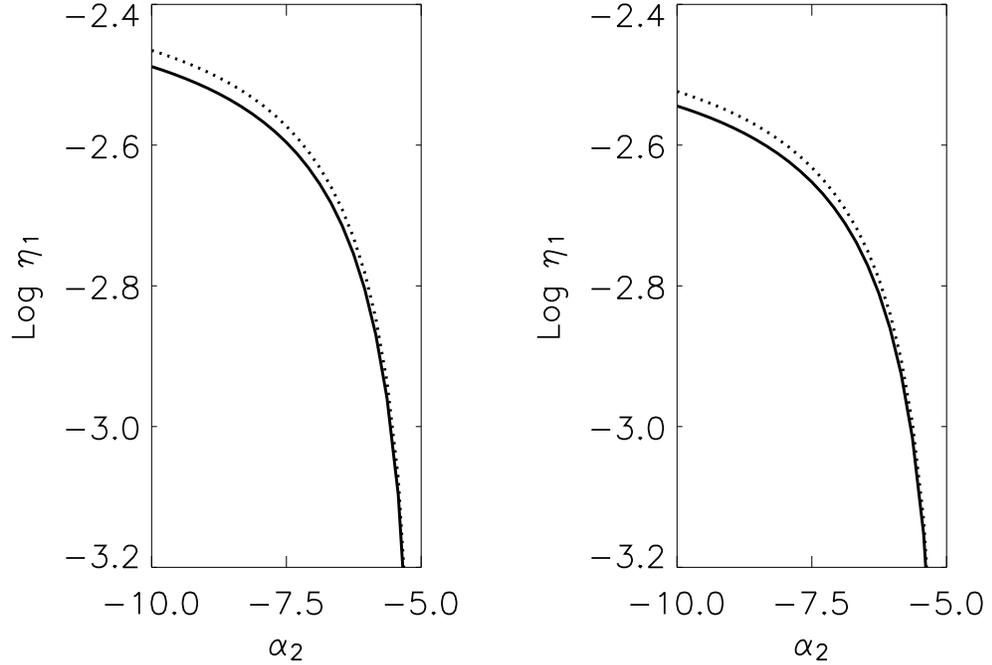}
\label{Fig 5}
\caption{Minimum efficiency of the first stage of the noise storm
emission process (accelerated electrons -- Langmuir wave production),
$\eta_1 \equiv L_{\rm L}/L_{\rm in}$, plotted as a function of
$\alpha_2$ using Equation~(\ref{neweq31}) with $\xi_c = \ximax$. Left
panel: $T_e = 10^6\,$K. Right panel: $T_e = 2 \times 10^6\,$K. Solid
line: $\nu = 169\,$MHz. Dotted line: $\nu = 327\,$MHz.}
\end{figure}

The efficiency of the first stage of the noise storm emission process is
defined by
\begin{equation}
\eta_1 \equiv {L_{\rm L} \over L_{\rm in}} \ ,
\label{neweq31}
\end{equation}
which can be evaluated by combining Equations~(\ref{neweq25}) and
(\ref{neweq28}). In Figure~5, we plot $\eta_1$ as a function of
$\alpha_2$ with $\xi_c = \ximax$ for the same two values of the coronal
temperature $T_e$ and observing frequency $\nu$ used in Figure~1. We
find that $L_{\rm L}$ decreases more rapidly than $L_{\rm in}$ as a
function of $\xi_c$, and consequently the results obtained by setting
$\xi_c = \ximax$ represent the {\it minimum} value of $\eta_1$
consistent with the required Langmuir wave intensity. We find that this
minimum value is quite insensitive to both the observing frequency and the
coronal temperature.

\begin{figure}
\includegraphics[width=5.5in]{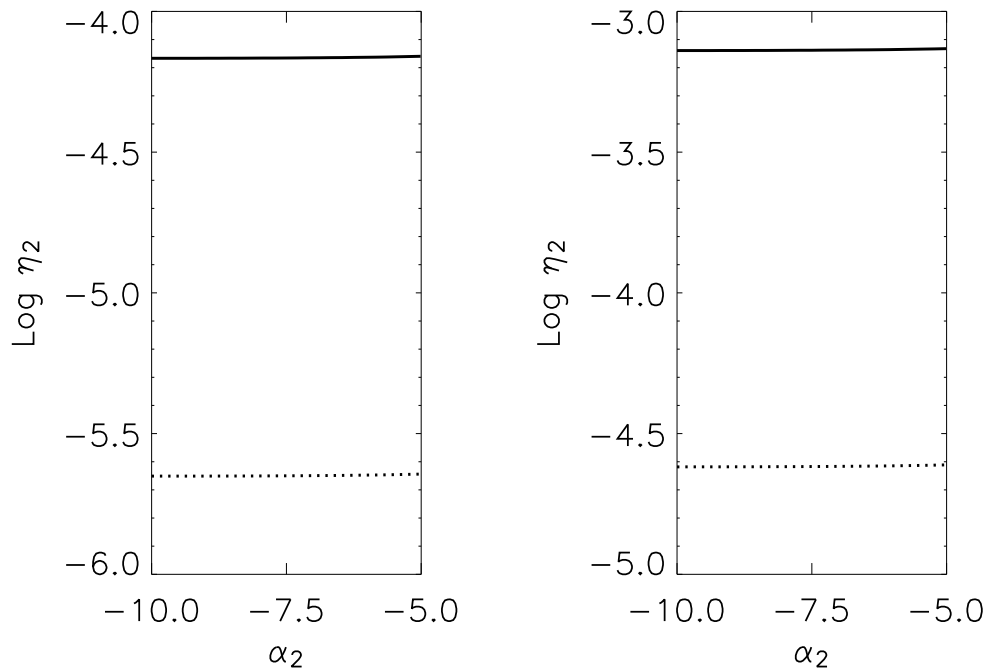}
\label{Fig 6}
\caption{Maximum efficiency of the second stage of the noise storm
process (Langmuir waves -- observable emission), $\eta_2 \equiv L_{\rm
out}/L_{\rm L}$, plotted as a function of $\alpha_2$ using
Equation~(\ref{neweq32}) with $\xi_c = \ximax$. Left panel: $T_e =
10^6\,$K. Right panel: $T_e = 2 \times 10^6\,$K. Solid line: $\nu =
169\,$MHz. Dotted line: $\nu = 327\,$MHz.}
\end{figure}

Our knowledge of the overall efficiency $\eta$ and the first-stage
efficiency $\eta_1$ can be combined to obtain an estimate of the
efficiency of the {\it second stage} of the noise storm generation
process, comprising the conversion of the Langmuir waves into the
observed noise storm radiation. The efficiency of the second stage is
defined by
\begin{equation}
\eta_2 \equiv {L_{\rm out} \over L_{\rm L}} \ ,
\label{neweq32}
\end{equation}
which is related to $\eta$ and $\eta_1$ via
\begin{equation}
\eta = \eta_1 \, \eta_2 \ .
\label{neweq33}
\end{equation}
In Figure~6 we plot the second-stage efficiency $\eta_2$ as a function
of $\alpha_2$ for the same two values of the coronal temperature $T_e$
and observing frequency $\nu$ used in Figure~1, with $\xi_c = \ximax$.
The result obtained is the {\it maximum value} for $\eta_2$. In general,
we find that $\eta_2 \lesssim 8 \times 10^{-4}$.

\subsection{Energy in the Langmuir wave population}

The energy density in the Langmuir wave population is an important
ingredient in the noise storm radiation process. Benz \& Wentzel (1981),
Jaeggi \& Benz (1982), and Benz (1982) have derived upper limits for this
quantity using radar measurements, observations of harmonic emission, and
polarization data. The general consensus is that $W_{\rm L} \lesssim
10^{-7} \, n_e\,k_{\rm B}\,T_e$ in the noise storm continuum sources.

\begin{figure}
\includegraphics[width=5.5in]{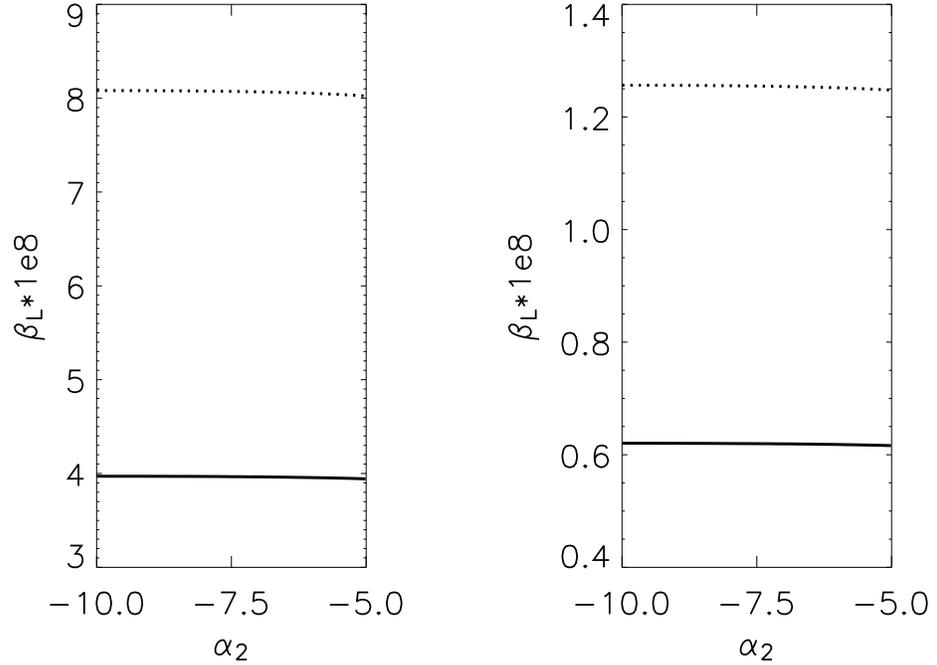}
\label{Fig 7}
\caption{Minimum value of the Langmuir wave energy density ratio,
$\beta_{\rm L} \equiv W_{\rm L}/(n_e\,k_{\rm B}\,T_e)$, plotted as a
function of $\alpha_2$ for $\xi_c = \ximax$. Left panel: $T_e = 10^6\,$K.
Right panel: $T_e = 2 \times 10^6\,$K. Solid line: $\nu = 169\,$MHz.
Dotted line: $\nu = 327\,$MHz.}
\end{figure}

By setting $\xi_c = \ximax$ and using Equation~(\ref{neweq29}), we can
obtain the lower limit on the Langmuir wave energy density $W_{\rm L}$
predicted by our theoretical model. In Figure~7 we plot the ratio
$\beta_{\rm L} \equiv W_{\rm L} /(n_e\,k_{\rm B}\,T_e)$ as a function of
$\alpha_2$ for the same two values of the coronal temperature $T_e$ and
observing frequency $\nu$ used in Figure~1. Since we have set $\xi_c =
\ximax$, it follows that this plot indicates the {\it minimum value} of
$\beta_{\rm L}$ consistent with the required intensity of the Langmuir
waves. We find that $8 \times 10^{-8} \gtrsim \beta_{\rm L} \gtrsim 6
\times 10^{-9}$, in agreement with the observational constraints derived
by Benz \& Wentzel (1981), Jaeggi \& Benz (1982), and Benz (1982).

\section{Summary and Discussion}

In this paper, we have focused closely on the first stage of the noise
storm emission process, namely the generation of a sufficiently intense
Langmuir wave distribution by the nonthermal electrons. This requires
that the nonthermal electron density fraction, $n_*/n_e$, exceeds the
minimum value given by the right-hand side of Equation~(\ref{neweq15}).
When this condition is satisfied, the damping of the Langmuir waves with
$k \approx k_c$ is dominated by the nonthermal electrons, which is an
essential ingredient for the production of bright noise storm emission
(Melrose, 1975; Robinson, 1978; Thejappa, 1991).

The main contribution of this paper lies in the fact that we have
self-consistently combined information about the minimum value of
$n_*/n_e$ with details of the electron acceleration process that were
derived in paper~1. In doing so, we have obtained limiting values for the
efficiency of the overall noise storm emission process, $\eta \equiv
L_{\rm out}/L_{\rm in}$, and also for the first-stage efficiency,
$\eta_1 \equiv L_{\rm L}/L_{\rm in}$, that are described by the single
parameter $\alpha_2$, which is the high-energy power-law index of the
nonthermal electron distribution (see Figures~4 and 5). In particular,
our theoretical results imply that the overall efficiency is bounded by
the upper limit $\eta \lesssim 10^{-6}$. We have also estimated
the second-stage efficiency, $\eta_2 \equiv L_{\rm out}/L_{\rm L}$,
describing the conversion of the Langmuir waves into observable noise
storm radiation, and conclude that the upper limit for this quantity
is $\eta_2 \lesssim 8 \times 10^{-4}$.

Our treatment also yields a lower limit for the dimensionless Langmuir
wave energy density in noise storm continuum sources, $\beta_{\rm L}
\equiv W_{\rm L}/(n_e\,k_{\rm B}\,T)$. We find that this lower limit
lies in the range $8 \times 10^{-8} \gtrsim \beta_{\rm L}
\gtrsim 6 \times 10^{-9}$. This interesting new theoretical estimate for
$\beta_{\rm L}$, a central parameter in the noise storm emission
process, is shown to be consistent with the observational data. The
efficiency estimates obtained here therefore represent an improved
understanding of the detailed energy budget for the noise storm
generation process.

It is interesting to contrast our approach with that taken by Thejappa
(1991). The model considered here includes an explicit treatment of the
acceleration of an isotropic, nonthermal electron distribution based on
the associated Fokker-Planck transport equation analyzed previously in
paper~1. Conversely, Thejappa's work focused on the emission of Langmuir
waves by an anisotropic loss-cone distribution of electrons in the
corona, and he did not consider the acceleration of the electrons in the
turbulent plasma. It is therefore suggestive to note that the values we
obtain for the nonthermal electron fraction $n_*/n_e$ are relatively
close to Thejappa's results. In the interpretation of Thejappa, it is
the {\it angular} anisotropy of the electrons in physical space (i.e.,
the loss-cone distribution) that provides the basic source of free
energy for the production of the Langmuir waves. However, our model
utilizes the anisotropy in {\it energy} space (i.e., the gap
distribution) as the source of free energy.

In reality, one would expect both types of anisotropies to coexist, and
therefore our results are in some sense complementary to Thejappa's.
Hence it is important to develop observational diagnostics for both
models that can be used to reveal the true nature of the underlying
nonthermal electron distribution as a function of both energy and
direction. It may be possible to develop the required observational
tests by making a detailed analysis of the corresponding predictions for
the time-dependent behavior of the associated radio signatures. We plan
to pursue this question in future work.

\acknowledgements

The authors are grateful to the anonymous referee for several
useful suggestions that helped to improve the paper. PS also
acknowledges the hospitality provided by Jagannath Institute of
Technology and Management, where part of this work was carried out.

\end{article}
\end{document}